\documentstyle[tighten,floats,preprint,prd,aps]{revtex}

\newcommand{\beq}{\begin{equation}}
\newcommand{\eeq}{\end{equation}}

\def\lap{\lower.5ex\hbox{$\; \buildrel < \over \sim \;$}}
\def\gap{\lower.5ex\hbox{$\; \buildrel > \over \sim \;$}}

\input epsf
\begin{document}

\title{Eternal inflation, black holes, and the future of civilizations.}
\author{J. Garriga\/$^{1,3}$, V.F. Mukhanov\/$^2$, K.D. Olum\/$^3$ and 
A. Vilenkin\/$^3$}
\address{
$^1$ IFAE, Departament de Fisica, Universitat Autonoma de Barcelona,\\
08193 Bellaterra (Barcelona), Spain\\
$^2$ Ludwig Maximilians Universit\"at, Sektion Physik\\
M\"unchen, Germany\\
$^3$ Institute of Cosmology, Department of Physics and Astronomy,\\
Tufts University, Medford, MA 02155, USA}

\maketitle

\begin{abstract}
We discuss the large-scale structure of the universe in inflationary
cosmology and the implications that it may have for the long-term
future of civilizations.  Although each civilization is doomed to
perish, it may be possible to transmit its accumulated knowledge to
future civilizations. We consider several scenarios of this sort.  If
the cosmological constant is positive, it eventually dominates the
universe and bubbles of inflationary phase begin to nucleate at a
constant rate.  Thermalized regions inside these inflating bubbles
will give rise to new galaxies and civilizations. It is possible in
principle to send a message to one of them.  It might even be possible
to send a device whose purpose is to recreate an approximation of the
original civilization in the new region.  However, the message or
device will almost certainly be intercepted by black holes, which
nucleate at a much higher rate than inflating bubbles.  Formation of
new inflating regions can also be triggered by gravitational collapse,
but again the probability is low, and the number of attempts required
for a positive outcome is enormous.  The probability can be higher if
the energy scale of inflation is closer to the Planck scale, but a
high energy scale produces a tight bound on the amount of information
that can be transmitted.  One can try to avoid quantum tunneling
altogether, but this requires a violation of quantum inequalities
which constrain the magnitude of negative energy densities.  However,
the limits of validity of quantum inequalities are not clear, and
future research may show that the required violation is in fact
possible. Therein lies the hope for the future of civilizations.

\end{abstract}

\section{eternal inflation}

Inflation is a period of accelerated expansion in the early universe.  
It is the only cosmological scenario that we have that can explain the 
large-scale homogeneity and flatness of the universe. During inflation 
the universe is expanded by a huge factor, so that we can see only a 
part of it, which is nearly homogeneous and flat.
The inflationary expansion is driven 
by the potential $V(\phi)$ of a scalar field $\phi$, which is 
called the inflaton. In Fig.\ 1 the inflaton is represented by a little 
ball that rolls down the potential hill. Near the top of the potential 
the slope is very small, so the roll is slow and $V(\varphi)\approx const.$ 
In this 
regime the universe expands exponentially,  
\beq
a(t)\propto e^{Ht}, 
\eeq 
with the expansion rate determined by the height of the potential, 
\beq
H^2={8\pi\over 3} V(\varphi).
\eeq
(Throughout the paper we use natural units in which 
$\hbar=c=G=1$)  When $\varphi$ gets to the 
steep part of the potential, it starts oscillating about the minimum, 
and its energy gets dumped into relativistic particles. The particles 
quickly thermalize at a high temperature, and the following evolution is
along the lines of the standard hot cosmological model. Thus, 
thermalization plays the role of the big bang in the inflationary scenario, 
and inflation prepares the initial conditions for the big bang (for a review
of inflation, see \cite{Lindebook}).

\begin{figure}[t]
\centering
\hspace*{-4mm}
\leavevmode\epsfysize=7 cm \epsfbox{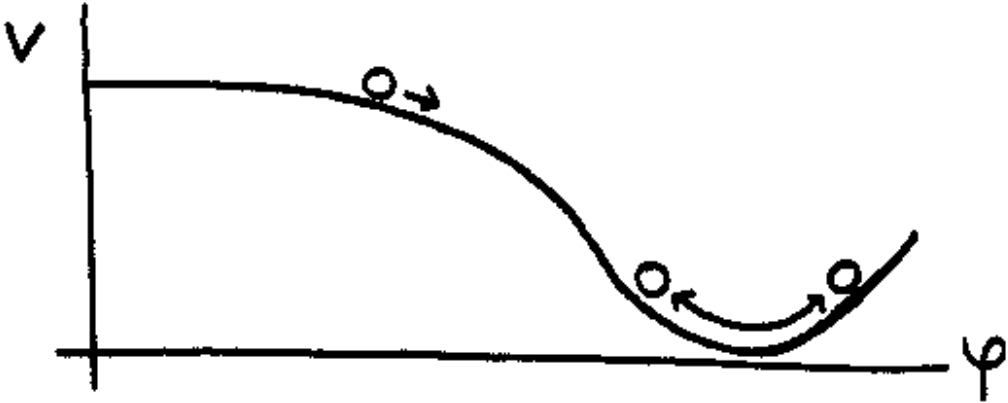}\\[3mm]
\caption[fig1]{\label{fig1}The inflaton potential}
\end{figure}

We hope that the shape of the potential $V(\varphi)$ will be determined from
the theory of elementary particles, but our present understanding of particle
physics is not sufficient for this task. Fortunately, some general features 
of inflation can be studied without a detailed knowledge of $V(\varphi)$.

One of the remarkable features of inflation is its eternal character.
This is due to quantum fluctuations of the inflaton. On the flat portion
of the potential, the force that drives $\varphi$ down the hill is small and
quantum fluctuations are important. The physics of the fluctuations is 
determined by the expansion rate $H$: the field $\varphi$ experiences quantum
jumps of magnitude $\delta\varphi\sim \pm H/2\pi$ on a time scale 
$\delta t\sim H^{-1}$. These jumps are not homogeneous 
in space: quantum fluctuations of $\varphi$ are not correlated over distances
larger than $H^{-1}$. In other words, fluctuations in non-overlapping regions 
of size $\sim H^{-1}$ are independent of one another.

Since quantum fluctuations of $\varphi$ are different at different locations, 
the inflaton does not get to the themalization point at the bottom of the
hill simultaneously everywhere in space. In some rare regions, the             
fluctuations will keep $\varphi$ at high values of the potential for much 
longer than it would otherwise stay there. Such regions will be ``rewarded'' 
by a large amount of expansion at a high rate $H$. The dynamics of the total 
volume of inflating regions in the universe is thus determined by two 
competing processes: the loss of inflating volume due to thermalization 
and generation of new volume at the high rate sustained by quantum 
fluctuations. Analysis shows that the second process ``wins'' for a 
generic inflaton potential $V(\varphi)$, and the total inflating volume grows 
with time \cite{AV83,Linde86,Starobinsky86}. 
Thus, inflation never ends completely: at any time there are 
parts of the universe that are still inflating. The spatial distribution 
of inflating and thermalized regions in an eternally inflating universe is
illustrated in Fig.\ 2. It was obtained in a numerical simulation
\cite{vitaly} for 
a double-well potential $V(\varphi)=\lambda(\varphi^2-\eta^2)^2$. 
Inflating regions are white, 
and the two types 
of thermalized regions corresponding to the two minima at $\varphi=\pm\eta$ 
are shown 
with different shades of grey. Different types of thermalized regions 
will generally have different physical properties.

\begin{figure}[t]
\centering
\hspace*{-4mm}
\leavevmode\epsfysize=10 cm \epsfbox{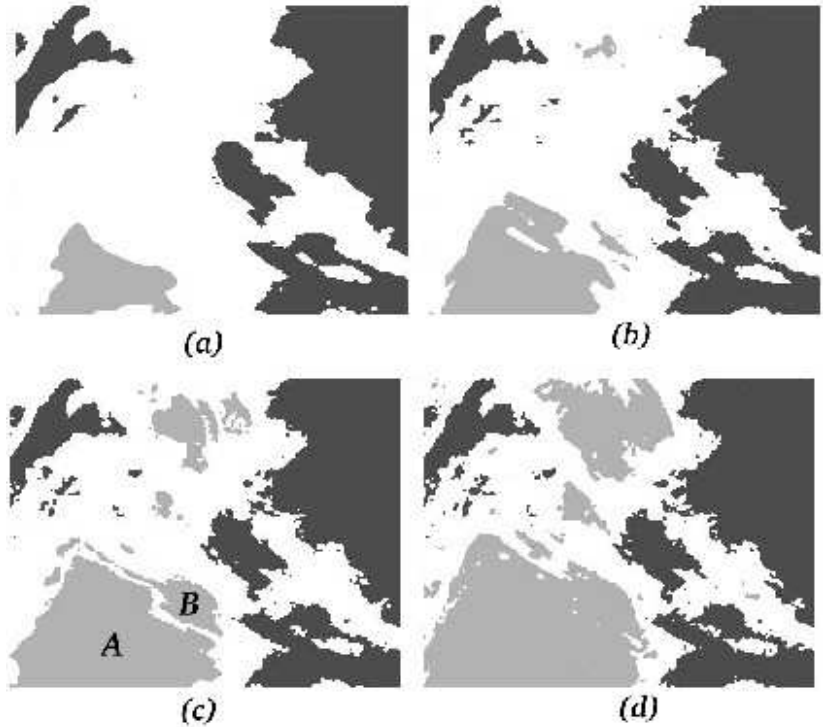}\\[3mm]
\caption[fig2]{\label{fig2}A simulation for the double well
model at four consecutive moments of time. Inflating regions are white, 
while thermalized regions with $\varphi=+\eta$ and $\varphi=-\eta$
are shown with different shades of grey \cite{vitaly}.}
\end{figure}

The spacetime structure of the universe in this model is illustrated 
in Fig.\ 3 using the same shading code. Now, the vertical axis is time 
and the horizontal axis is one of the spatial directions. The boundaries 
between inflating and thermalized regions play the role of the big bang 
for the corresponding thermalized regions. In the figure, these boundaries 
become nearly vertical at late times so that it appears that they 
correspond to a certain position in space rather than to a certain moment 
of time. The reason is that the horizontal axis in Fig.\ 3. is the co-moving 
distance, with the expansion of the universe factored out. The physical 
distance is obtained by multiplying by the expansion factor $a(t)$, which 
grows exponentially as we go up along the time axis. If we used the 
physical distance in the figure, the themalization boundaries would 
``open up'' and become nearly horizontal (but then it would be difficult 
to fit more than one thermalized region in the figure).

\begin{figure}[t]
\centering
\hspace*{-4mm}
\leavevmode\epsfysize=10 cm \epsfbox{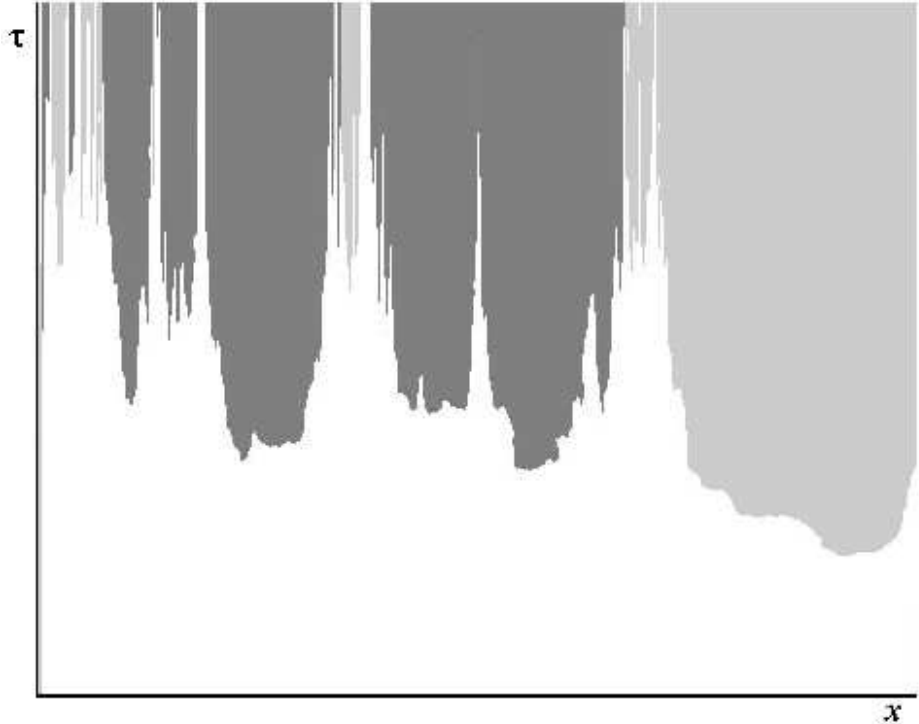}\\[3mm]
\caption[fig3]{\label{fig3} Spacetime structure simulation for 
the double-well model. Inflating regions are white, 
while thermalized regions with $\varphi=+\eta$ and $\varphi=-\eta$
are shown with different shades of grey \cite{vitaly}.}
\end{figure}

The spacetime structure of a thermalized region near the thermalization 
boundary is illustrated in Fig.\ 4.
The thermalization is followed by a hot radiation era       
and then by a matter-dominated era during which luminous galaxies are       
formed and civilizations flourish. All stars eventually die, and 
thermalized regions become dark, cold and probably not suitable for life 
(see the next 
section). Hence, civilizations are to be found within a layer of finite 
(temporal) width along the thermalization boundaries in Fig.\ 3.

\begin{figure}[t]
\centering
\hspace*{-4mm}
\leavevmode\epsfysize=8 cm \epsfbox{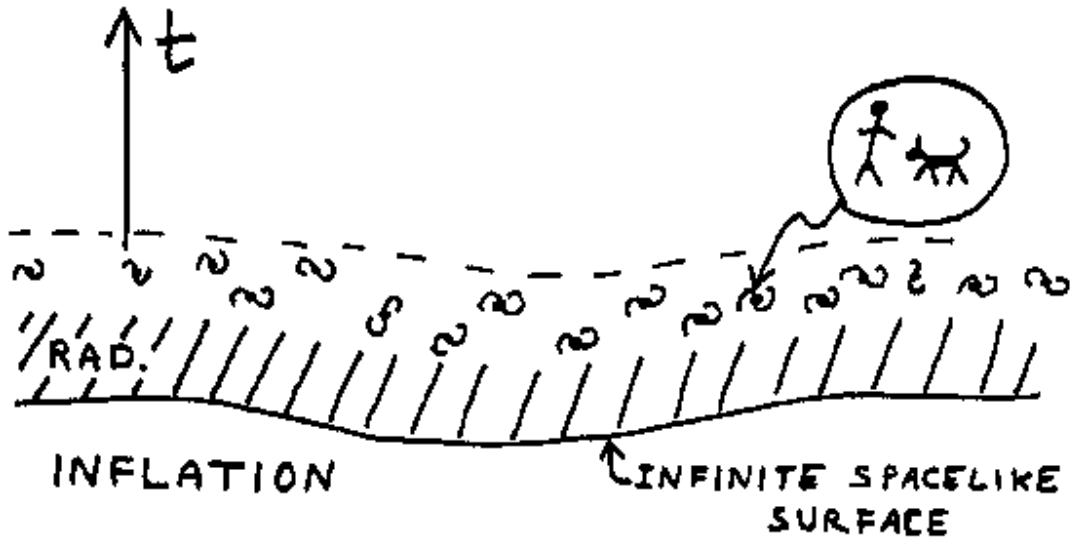}\\[3mm]
\caption[fig4]{\label{fig4} Spacetime structure of a thermalized region
near the thermalization boundary.}
\end{figure}

For an observer inside one of the thermalized regions, the thermalization 
boundary (the ``big bang'') is in his past, so he cannot reach this boundary, 
no matter how fast he moves 
\footnote{Thermalization boundaries are infinite 
spacelike hypersurfaces. With an appropriate choice of the time coordinate, 
each thermalized region is an infinite sub-universe containing an infinite 
number of galaxies.}.
Even at the speed of light, it is impossible 
to send a signal that would cross the boundary and get into the inflating 
region. It follows that different thermalized regions are causally 
disconnected from one another: it is impossible to send signals between 
different regions, and the course of events in one region can be in no way 
affected by what is happening in another. 

This conclusion may be avoided in the presence of a peculiar form of
matter described by a field equation with a special non-linear
gradient term in which sound propagates faster than the speed of light
\cite{GM99}. The existence of such matter is consistent with the
principles of the theory of relativity: the Lagrangian is perfectly
Lorentz invariant, but the cosmological solutions select a preferred
frame in which the speed of sound can be superluminal.  In what
follows, however, we shall disregard this somewhat exotic possibility.
Note also that in order to get from one thermalized region to another,
the sound waves have to cross the inflating region that separates
them.  As a result, unless the speed of sound is so large that the
trip can be made almost instantaneously, the wavelength may get
stretched by an enormous expansion factor. In this case the period of
the waves would become too long for even one oscillation during the
lifetime of a star in the target region, and it is hard to see how
such waves could transmit any information.

\section{Messages to the future}

Let us now consider the future prospects for a civilization on a cosmological 
timescale. It appears very unlikely that a civilization can survive forever. 
Even if it avoids natural catastrophes and self-destruction, it will in the 
end run out of energy. The stars will eventually die, and other sources of 
energy (such as tidal forces) will also come to an end. Dyson 
\cite{Dyson} has argued 
that civilizations may still survive indefinitely into this cold and dark 
future of the universe by constantly reducing the rate of their energy 
consumption. This must be accompanied by a corresponding reduction in the 
rate of information processing. In the asymptotic future both rates become 
infinitesimal, but Dyson argued that the total amount of information 
processed by a civilization may still be infinite. The ``subjective'' 
lifetime of such a civilization would then also be infinite, as well as its 
physical lifetime.  Nevertheless, a more detailed analysis of the 
problem has been recently given by Krauss and Starkman \cite{kraussstarkman}, 
with the conclusion that the steady decrease in the rate of information
processing proposed by Dyson seems physically impossible to achieve. Thus,  
it appears that an eternal civilization is impossible, even in principle.

If we are doomed to perish, then perhaps we could send messages, or
even representatives, to future civilizations?  Those civilizations
could also send messages to the future, and so on. We would then
become a branch in an infinite ``tree'' of civilizations, and our
accumulated wisdom would not be completely lost.  Here we shall
consider the feasibility of some scenarios of this sort
\cite{Lindenote}.

\subsection{Recycling universe}

In an eternally inflating universe, new thermalized regions will continue to 
be formed in the arbitrarily distant future. These regions will go through 
radiation and matter-dominated periods, will form galaxies and evolve 
civilizations. However, as we discussed at the end of the preceding section, 
different thermalized regions are causally disconnected from one another 
and communication between them is impossible. We can of course send messages 
to other civilizations within our own thermalized region. But the resulting 
tree of civilizations would necessarily be finite, and its time span would 
be bounded by the same energetic factors that limit the lifetime of a single 
civilization. 

The spacetime structure of the universe illustrated in Fig.\ 3 assumes that
the vacuum energy density, otherwise known as the cosmological constant
$\Lambda$, is equal to zero. However, observations of distant supernovae 
performed by two independent groups during the last year suggest that the 
universe is 
now expanding with an acceleration, indicating that $\Lambda$ 
has a small positive value
\cite{SN}. A nonzero $\Lambda$, no matter 
how small, changes the large-scale structure of
the universe \cite{recycling}. As the universe expands, the density of matter 
goes down, $\rho_m\propto a^{-3}$, while the vacuum energy density remains 
constant, so eventually
the universe gets dominated by the cosmological constant. After that it
expands exponentially, as in Eq.\ (1) , but at a very small rate 
$H_{\Lambda}=(\Lambda/3)^{1/2}$.

In a universe with $\Lambda >0$, there is a finite constant probability 
for the inflaton field to tunnel quantum-mechanically from its vacuum 
value, where $V(\varphi)=\Lambda/8\pi$, to the values in the 
inflationary range near the top of the 
potential. The tunneling occurs within a spherical volume of radius 
$\sim H^{-1}_{\Lambda}$
(a ``bubble''), with a probability per unit volume per unit time
${\cal P}\propto \exp(S_b-3\pi\Lambda^{-1})$ 
\cite{leeweinberg}. Here $S_b$ is the action of the instanton solution 
responsible for the tunneling. (Note that $\cal{P}$ vanishes for $\Lambda=0$). 
Each inflating bubble develops into a 
fully-fledged eternally inflating region of the universe. In the course
of its evolution, it forms an infinite number of thermalized regions, each
containing an infinite number of galaxies. These thermalized regions get
later dominated by the cosmological constant, with subsequent nucleation of
new inflationary bubbles, and so on \cite{recycling}.
The large-scale structure of such a ``recycling'' universe is illustrated 
in Fig.\ 5.

\begin{figure}[t]
\centering
\hspace*{-4mm}
\leavevmode\epsfysize=10 cm \epsfbox{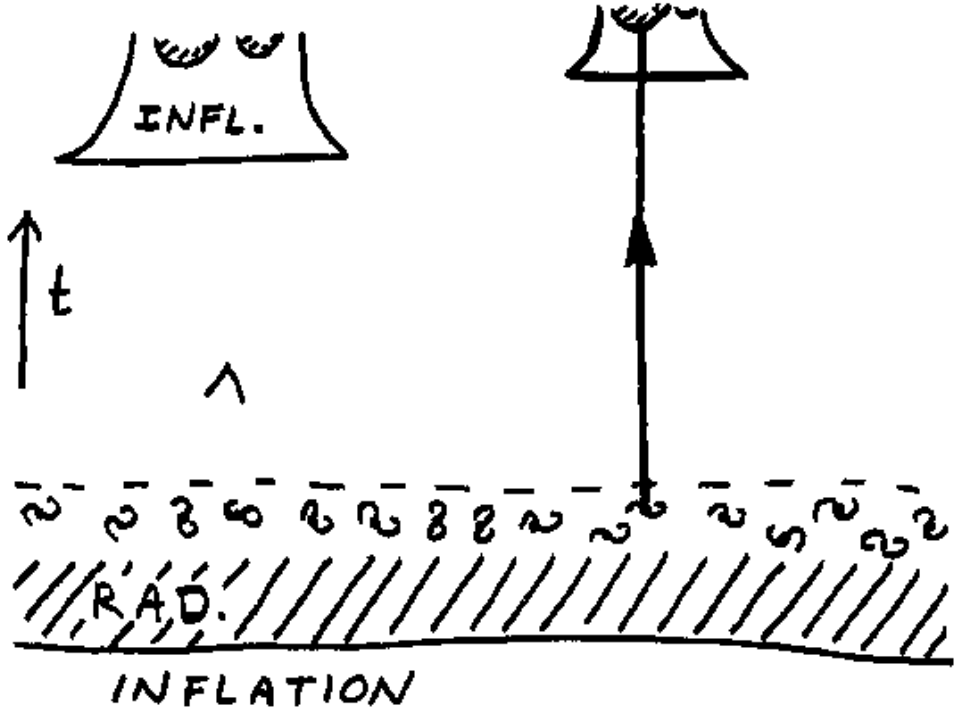}\\[3mm]
\caption[fig5]{\label{fig5} Global structure of a recycling 
universe. After the universe has thermalized, the cosmological
constant $\Lambda$ drives an accelerated expansion with constant
Hubble rate. This triggers the nucleation of ``bubbles'' of the
new phase at a constant rate. Two of these bubbles are represented
in the higher part of the figure. In a ``recycling'' universe, a message
whose worldline is indicated by a vertical arrow can travel from 
one civilization in the thermalized region at the bottom of the
figure to another civilization which resides in a future thermalized 
region.  Most messages, however, are intercepted by nucleating black
holes.} 
\end{figure}

The recycling nature of the universe opens the possibility of sending a 
message to a future civilization. All one needs is a very strong and durable 
container. One simply puts the message in the container and sends it into 
space. In due course, the universe gets dominated by the cosmological 
constant, and inflating bubbles begin to nucleate.  The hope is that 
the container will be engulfed by one of the bubbles. 
%\footnote{The probability for the container not to 
%be hit by a bubble for a time t decreases exponentially with time.}

The problem with this scenario is that inflating bubbles are not the
only things that can nucleate in a $\Lambda$-dominated universe.  
There is also a constant rate of nucleation of black holes, ${\cal
P}_{bh} \propto \exp(-\pi/\Lambda)$.  For the value of $\Lambda\sim
10^{-122}$ suggested by observations, this is greater than the rate of
bubble nucleation by a factor $\sim \exp(10^{122})$.  Thus, the
message-carrying container will almost certainly be swallowed by a
black hole.  In order to beat the odds, one would have to send more than
$\sim \exp(10^{122})$ containers.

\subsection{Black holes and an upper bound on information}
\label{sec:bits}

Instead of relying on bubble nucleation, one can take a more active
approach\footnote{This possibility was suggested to us by
E. Guendelman.}.  Quantum nucleation of an inflating region can be
triggered by a gravitational collapse in our part of the universe
\cite{guendelman} (see Fig.\ 6).  
One has to generate an implosion of a small
high-energy vacuum region, leading to gravitational collapse, with a
message-carrying container at the implosion center. All one sees is
the formation of a black hole; a new inflating region may or may not
be inside it. The tunneling probability for the formation of such a
region is \cite{guendelman} \beq {\cal P}\propto \exp(-C/H^2),
\label{pbh} 
\eeq
where $H$ is the
expansion rate in the new inflating region and $C\sim 1$ is a constant
coefficient.  For a grand-unification-scale inflation, $H\sim 10^{-7}$
and ${\cal P}\sim\exp(-10^{14})$.

\begin{figure}[t]
\centering
\hspace*{-4mm}
\leavevmode\epsfysize=10 cm \epsfbox{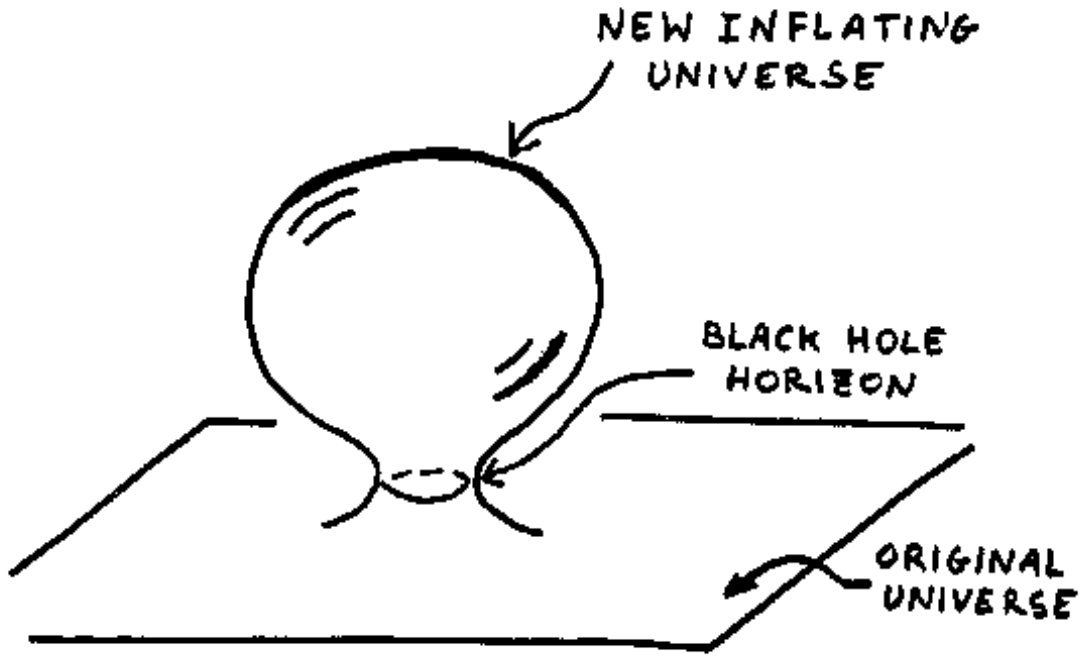}\\[3mm]
\caption[fig6]{\label{fig6} A new inflating region can be triggered
by gravitational collapse in our part of the universe. The figure
represents a space-like section of the resulting space-time. The inflating
region eventually ``pinches off'' the original universe.} 
\end{figure}

The exponential suppression of the tunneling disappears if inflation
is at the Planck scale, $H\sim 1$.  But then we encounter a different
problem: there seems to be an upper bound on the amount of information
that can be sent.  Indeed, if the container is to survive a period of
inflation at a high expansion rate $H$, then its diameter must be
smaller than the horizon, $D<H^{-1}$, or else the container would be
torn apart by the expansion.  (One could imagine spreading the
information among multiple containers, but it would be very difficult
for it to be reassembled after the containers had been separated by
vast distances during inflation.)  Now, it has been argued that the
amount of information contained in a sphere of diameter $D$ is bounded
by \cite{holographic}
\begin{equation}
I\leq A/4,
\label{info}
\end{equation}
where $A=\pi D^2$ is the surface area of the sphere (in natural units).
The general validity of this ``holographic'' bound
is still a matter of debate. However, for our particular case,
the inequality (\ref{info}) can be derived directly from the generalized 
second law of thermodynamics.
The maximum information that a package can contain at a given value of 
its energy is equal to the logarithm of the number of microstates of
this package compatible with the given energy. That is, the maximum
amount of information coincides with the thermodynamical entropy of 
the package in thermal equilibrium. Now, we can think of a process
in which the package collapses into a black hole (or is swallowed 
by a very small black hole). The
entropy of the package has to be less than the entropy of the 
resulting black hole, which is equal to one fourth of its surface area
\cite{beckenstein}.
Since the largest black hole that can exist in an inflating universe
has radius $1/\sqrt{3}H$ \cite{nariai},
this implies that the largest amount of information that can be sent is 
\beq
I_{max}=1/12H^2\,.
\label{imax}
\eeq
The usual values of $H$ in inflationary models range from $10^{-7}$ for
grand unification scale inflation to $10^{-34}$ for electroweak scale
inflation, yielding $I_{\text{max}}\sim 10^{13} - 10^{68}$.  This can
be compared with $I\sim 10^{10}$ for the human genome, $I\sim 10^7$
for a typical book, and $I\sim 10^{15}$ for all books in the Library
of Congress.  Even if one makes no assumptions about the model, with
$I\gtrsim 10^{7}$, Eqs.\ (\ref{imax}),(\ref{pbh})
require $H^2\lesssim 10^{-8}$ and ${\cal P}\lesssim \exp(-10^8)$.
This is an improvement over the case of nucleating bubbles, but still
the number of attempts required to beat the odds far exceeds the
number of elementary particles in the visible part of the universe
\cite{Lindenote}.

\subsection{Negative energies}

The root of the problem appears to be in the tiny tunneling
probability (\ref{pbh}), so it is reasonable to inquire whether or not
a new inflating region can be created without quantum tunneling.  This
question was addressed by Farhi and Guth \cite{Farhi} who concluded
that the answer is ``no'', provided that a few very general
assumptions are satisfied.  Among these  assumptions the most
important  is the weak energy
condition, asserting that the energy density measured by any observer
is never negative.  Although it is satisfied in all familiar states of
matter, this condition is known to be violated in certain
 states of quantum fields (e.g., the electromagnetic field or scalar
fields that are used in inflationary models).  

The newly-created inflating region should have an extent $\gtrsim
H^{-1}$, where $H$ is the inflationary expansion rate.  Hence, one
needs to violate the weak energy condition in a spacetime region
\beq
\Delta L\sim \Delta t \gtrsim H^{-1}.  
\label{neg1}
\eeq
The required magnitude of the
negative energy density is 
\beq
-\rho \gtrsim H^{-2}.
\label{neg2}
\eeq

For non-interacting fields, the magnitude and the duration of 
violations of the weak energy condition are constrained by the
so-called quantum inequalities \cite{Ford},   
\beq
-\rho\lesssim (\Delta t)^{-4}.
\label{neg3}
\eeq
Combining this with Eqs.\ (\ref{neg1}),(\ref{neg2}), we see that all
conditions can be satisfied only for super-Planckian inflation with
$H\gtrsim 1$.  Then again, one has  to face the information bound
(\ref{imax}).  

The negative energy density in Eq.\ (\ref{neg3}) should be understood in
the sense of a quantum expectation value.  There are quantum
fluctuations about this value, and occasionally the fluctuation may
get large enough to provide the required negative energy in a
sufficiently large region.  But again, such fluctuations are
suppressed by an exponentially large factor \cite{Linde91}.

It should be noted, however, that the validity of quantum inequalities
like (\ref{neg3}) is not certain beyond the case of free quantum
fields for which they have been established \cite{Solomon}.  For
example, the Casimir energy density \cite{Casimir} of the
electromagnetic vacuum between two conducting plates appears to be
negative and permanent, in violation of (\ref{neg3}).

\subsection{Limiting curvature}

Quantum effects at high curvature could significantly modify the 
dynamics of the gravitational field.  In particular, it has been suggested
\cite{Slava} that there exists a limiting curvature $R_{max}$ which can 
never be exceeded.  This would result in a drastic change in the final
stages of the gravitational collapse.  It has been argued in \cite{Slava} 
that the black hole singularity and the adjacent high curvature region in
the black hole interior get replaced by a de Sitter space of the limiting 
curvature $R_{max}$.  In the Schwarzschild solution describing the usual
black hole, the spacetime could not be continued beyond the singularity,
but now the de Sitter space extends all the way into a new inflating 
universe which is in the absolute future with respect to the original one.
Assuming that the state of limiting curvature is metastable and decays
dumping its energy into particles, we will have formation of thermalized 
regions and the usual picture of eternal inflation inside the black hole.

In the course of the gravitational collapse, the effective energy-momentum
tensor in this model should develop negative energy densities that violate 
quantum inequalities.  In this sense, the limiting curvature conjecture
can be regarded as a specific example of a more general class of models
with negative energies.  An important difference, however, is that with a
limiting curvature, inflating universes automatically form inside black 
holes, with no effort required on our part.  To send an information 
container to another universe, all we need to do is to drop it into a black 
hole.  And in a recycling universe, black holes are no danger at all:
they simply provide a passage to a new inflating region.

The curvature of de Sitter space is $R=12 H^2$, and it follows from 
(\ref{imax}) that the largest amount of information that can be sent is
\beq
I_{max}=1/R_{max}.
\eeq
We thus see that a non-trivial amount of information can be sent only if 
$R_{max}$ is well below the Planck scale, $R_{max}\ll 1$.

\subsection{Summary}

To summarize, it appears that all mechanisms that involve quantum
tunneling are doomed to failure, because of extremely small tunneling
probabilities.  Creation of new inflating regions without quantum
tunneling requires a violation of the weak energy condition that is in
conflict with quantum inequalities.  It is not clear how seriously
this constraint is to be taken, since we don't know to what extent
quantum inequalities apply to interacting fields.  Since the future of
the civilization depends on the outcome, this can be regarded as a
good reason to increase funding for the negative energy research!  In
the following Section, we shall take an optimistic attitude and assume
that advanced civilizations will figure out how to get around quantum
inequalities.\footnote{In the absence of energy conditions, wormholes
\cite{VisserBook} would also be possible, and perhaps could be used to
communicate between different regions.  However, the maintenance of
a long-lasting wormhole requires negative energies to exist
indefinitely, whereas creating a new inflating region as above
requires them only for a short period of time.  We will not consider
wormhole scenarios any further here.}

\section{Discussion}

Suppose now that we have resolved all ``technical'' problems
associated with sending messages to future civilizations.  This
includes learning how to generate negative energy in a sufficiently
large volume, so that we can create new inflating regions (in case
such negative energies are not generated without our intervention, 
due to the limiting curvature or some other mechanism), and
designing containers for information that can survive the negative
energy density, a period of inflation, and the subsequent hot
radiation era.  Now, what should our strategy be?  How many containers
should we send?  Should we search for a message in our part of the
universe, in case it was created by an advanced civilization that
lived prior to our inflation?

Let us first address the question about the number of messages that we
should send.
A mission can be regarded a success if the information is successfully
transmitted to a civilization in the new region that
is capable of sending messages of its own.  If the probability of
success is $p$, then one should send $N\gg 1/p$ information packages,
to make sure that at least one of them succeeds.  The ultimate goal is
to initiate an infinite tree of civilizations, so that our
knowledge can propagate indefinitely into the future.
If ${\bar N}$ is the average number of missions launched by the
civilizations that form the tree, then the average number of
civilizations in each
successive generation is related to the number of their predecessors
by a factor $p{\bar N}$. Thus, for ${\bar N}>1/p$ there
is a non-zero probability that the process will never end and the
total number of civilizations in the tree will be infinite.

We next turn to the implications of the bound (\ref{imax}) on the
amount of information that can be contained in a message.  The maximal
information $I_{max}$ is equal to the logarithm of the total number
of quantum states available to the information package.  This means
that there is only a finite number of possible messages, $N\leq
\exp(I_{max})$.  There should, therefore, be an ``optimal'' message
which gives the highest rate of reproduction, by inducing its
recepients to send the largest number of successful missions.  This
optimal message will inevitably be discovered in each
infinite tree of civilizations and will eventually become the dominant
message, in the sense that the fraction of civilizations receiving
other messages within that tree will exponentially approach zero 
as we go up along the tree.
(Clearly, the ``optimal'' message should come with
the instruction that it should be passed along without change.)

How likely is it that our civilization will receive a message from the
past?  Let us assume that some finite fraction of the ``orphan''
civilizations, who do not receive messages from their ancestors,
succeed in initiating infinite trees.  Then, for each civilization that
does not receive a message, there is an infinite number of
civilizations (in its future) who do receive messages, with most of
the civilizations receiving the optimal message.  One may be inclined to
conclude that our civilization is most likely to be a recepient of the
optimal message.  One has to be careful, however, because the numbers
of civilizations that do and do not receive messages are both
infinite, and comparing infinite sets is a notoriously ambiguous task.  
The result crucially depends on how one maps one set onto the other.

Instead of comparing our civilization with its descendants in the
infinite tree, it appears more natural to compare it with other
civilizations in the same thermalized region.  We cannot prove that
this is the correct procedure, but we note that a similar approach
appears to work well for calculating probabilities in an eternally
inflating universe \cite{AV98,vitaly}.  
Our thermalized region contains an infinite number of
civilizations, but it can contain only a finite number of information
packages from our predecessors.  Hence, the probability for a package
to be anywhere in our neighborhood is zero.  If we do receive a
message (which of course is extremely unlikely), then the same
argument can applied to the civilization that sent it to us.  With a
100\% probability, that civilization was the first in line and received no
messages from its predecessors.  So it is very unlikely that the
message we receive is the optimal message.

Since the probablity of receiving a message is zero, our descendants
would be foolish to waste their resources to search for messages.
We could therefore try to make the information container very conspicuous.  It
could, for example, transmit its message in the form of
electromagnetic waves, using some star as a source of energy.  (Of
course, the container should then be programmed to search for a
suitable star.)  Perhaps a more reliable plan might be for the
``message'' to instead be a device which reproduces our civilization
in the new region, rather than waiting for new civilizations to
evolve.  In such case one can consider this process to be the
continuation of the old civilization, rather than a new civilization
at all.  One can even imagine some individual members of the old
civilization surviving in the container into the new region, perhaps
by having their physical form and state of knowledge encoded in some
compact and durable way for later reproduction.  However, due to the
limitations on
the amount of information that can be included in the container, 
it may be necessary to send ``simplified'' representatives if the energy scale
of inflation is high.

Since there is a finite number of possible messages, the process of
accumulation of knowlege will inevitably halt, and one can ask if 
there is any point in generating an infinite tree of civilizations.  
There may well
be a point at the beginning of the tree, 
while the limits on the size of the message are not yet reached.  
After the optimal message is hit upon, there may be 
no point, but the process is likely to continue anyway.

We are grateful to Eduardo Guendelman for a useful
discussion.  J.G., V.F.M. and A.V. are grateful to Edgard Gunzig and
Enric Verdaguer for their hospitality in Peyresq where part of this
work was completed.  This work was supported in part by CIRIT grant
1998BEAI400244 (J.G.), by a NATO grant CRG 951301 (J.G.),
and by the National Science Foundation (K.D.O. and A.V.).


\begin{thebibliography}{99}
\bibitem{Lindenote}
Linde \cite{Linde91} suggested 
that the message could be sent in the form of the
values of the fundamental constants of Nature.  However, it is not
clear that the constants in a newly created inflating region can be
different from those in the parent universe, and if so, whether or not
we can have any control over their values.  We shall not consider this
possibility in the present paper.
\end{thebibliography}
\end{document}